\documentstyle[12pt,twoside,fleqn,espcrc1,epsfig]{article}
\newcommand{\be}{\begin{equation}}
\newcommand{\ee}{\end{equation}}
\newcommand{\bea}{\begin{eqnarray}}
\newcommand{\eea}{\end{eqnarray}}
\newcommand{\dpf}{\displaystyle\frac}

\newcommand{\AmS}{{\protect\the\textfont2
  A\kern-.1667em\lower.5ex\hbox{M}\kern-.125emS}}

\title{$\rho$ meson broadening and dilepton production  
in heavy ion collisions\footnote
{Supported by AvH, BMBF, DFG and GSI, NSERC, FCAR}}

\author{S.\ Gao${}^{\rm a}$\footnote
{Alexander von Humboldt Fellow}, C.\ Gale${}^{\rm b}$, C. Ernst${}^{\rm a}$,
H.\ St\"ocker${}^{\rm a}$ and W.\ Greiner\address{
Institut f\"ur Theoretische Physik der J.W.
Goethe-Universit\"at\\
\ \ Postfach 111932, D-60054 Frankfurt a. M., Germany \\
$^{\rm b}$Physics Department, McGill University\\
\ \ 3600 University St., Montr\'eal, QC, H3A 2T8, Canada}}

\begin{document}
\maketitle
\begin{abstract}
The modification of the width of the rho meson due to in-medium
decays and collisions is evaluated. In high temperature and/or high density
hadronic matter, the collision width is much larger than the one-loop decay 
width. The large width of the $\rho$ meson in matter seems to be consistent 
with some current interpretations of the $e^+ e^-$ mass spectra measured at 
the CERN/SPS.
\end{abstract}

\section{INTRODUCTION}

The measurement and calculation of dilepton and photon emission
in heavy ion collisions represent great interest from a theoretical
point of view. Once the radiation is formed, it can propagate virtually
unscathed to the detectors, owing to the smallness of the fine structure
constant $\alpha$. This signal is also sensitive to the hot and dense 
phase of the nuclear reaction. This can only be probed indirectly by 
hadronic observable. Lepton pairs have  been proposed as valuable tools 
in studies of reaction dynamics and also as a possible signature of the 
quark-gluon plasma which is a prediction of QCD \cite{meor}. At SPS 
energies at CERN, the HELIOS \cite{hel98} collaboration has measured the 
low mass sector and more recently the CERES collaboration has
produced tantalizing experimental results for soft pairs \cite{drees98}. 
The intermediate mass sector has been covered by HELIOS, and by the NA38 
and NA50 collaborations. 

Our goal here is to compute dielectron emission from 
$\pi \rho \to \pi e^+e^-$ processes,
including $a_1$, $\pi$, $\rho$, $\omega$ mesons in the intermediate states.
We will highlight interference effects and also show the explicit
consequences of rho meson collision broadening in matter.

\section{DILEPTON PRODUCTION RATES}

At CERN energies, nucleus-nucleus reactions have been seen to create a
medium that is meson-dominated. Thus, a large number of lepton production 
calculation have concentrated exclusively on meson reactions. The two-body 
cases were studied  in Ref. \cite{gali}. There, reactions of the type 
$a + b \to e^+ e^-$ were calculated and their strength compared with each
other. The rates obtained from that study and from very similar ones were 
used to compare with CERES experimental results, using a variety of 
dynamical simulation models. Strikingly, all results turned out to be in 
agreement with each other, but in significant disagreement with the 
data \cite{drees98}. The situation called for additional physics 
ingredients. A first school of thought successfully fitted the CERES data 
by invoking a dropping $\rho$ meson mass reflecting a many-body effect, 
possibly a chiral phase transition precursor \cite{likobr}. A second school 
of thought successful in reproducing the dilepton data invokes a widening 
of the $\rho$ meson which originates mainly from its strong coupling to 
baryon resonances \cite{rapp}. While each of these two possibilities is
enticing, they still need to be disentangled and furthermore the
physical conditions necessary for each to occur may actually coexist.
In the spirit of a systematic study of this physics, it is necessary to 
explore the two-loop case. 

We consider a thermal medium of pions and their main resonance,
$\rho$ mesons. It has been shown that the dominating contribution to
lepton pair emission from reactions other than two-body is actually
$\pi \rho \to \pi e^+e^-$ \cite{bdr97,kh98}.  Our goal here is to revisit
this contribution and to extend the previous evaluations by {\bf (a)} 
including a more complete set of mesons mediating the
$\pi \rho$ interaction, {\bf (b)} providing a consistent calculation 
of $\rho$ collisional
broadening, and its influence on the lepton pair production rates.

The dilepton production rate for $ 1 + 2 \to 3 + l^+ l^-$ process can be
written as \cite{lich}:
\begin{eqnarray}
\dpf{dN}{d^4x} &=& {\cal N} \int{d^3p_1\over(2\pi)^3 2E_1} \,
         {d^3p_2\over(2\pi)^3 2E_2}\, {d^3p_3\over(2\pi)^3 2E_3}\,
         { n}_1(E_1) { n}_2(E_2) [1 + { n}_3(E_3)]\,  \nonumber \\
      & &\ \times \  {d^3p_+\over(2\pi)^3 2E_+}\, {d^3p_-\over(2\pi)^3 2E_-}
        (2 \pi)^4 \delta(p_1+p_2-p_3-p_+-p_-)\, 
       \cdot \overline{|{\cal M}_{12 \to 3 l^+l^-}|^2} \, ,
\end{eqnarray}
where ${\cal N}$ is an overall degeneracy factor, $n_i(E_i)$ are
statistical distribution functions, and ${\cal M}$ is the 
scattering amplitude. 

In order to calculate the dilepton production rates from
reaction $\pi \rho \to \pi e^+e^- $, the following four processes
should be considered:
(1) $\pi^{\pm} \rho^0 \to (\pi^{\pm} \, , a_1^{\pm})
        \to \pi^{\pm} e^+e^- \, ,$
(2) $\pi^0 \rho^{\pm} \to (\pi^{\pm} \, , a_1^{\pm} \, , \rho^{\pm})
        \to \pi^{\pm} e^+e^- \, , $
(3) $\pi^{\pm} \rho^{\mp} \to (\pi^{\pm} \, , \rho^{\pm} \, ,
                \omega) \to \pi^0 e^+e^-\, ,$
(4) $\pi^0 \rho^0 \to \omega \to \pi^{0} e^+e^- \,$.
     In the processes (1) - (3), the appropriate
$\pi \pi \rho \rho$ four point diagram is included.
It is well known that the numerical integration for 
$\pi \rho \to \pi \to \pi e^+e^-$ (t-channel) is singular, we
regulate it following the effective approach of Peierls\cite{pei}. 
We have checked that the results are gauge invariant in the 
electromagnetic sector. 

\section{$\rho$ COLLISION RATE IN THERMAL BACKGROUND}

\begin{figure}[hbt]
\begin{minipage}[t]{70mm}
\vspace{-6.2cm}
{\centerline{\epsfig{file=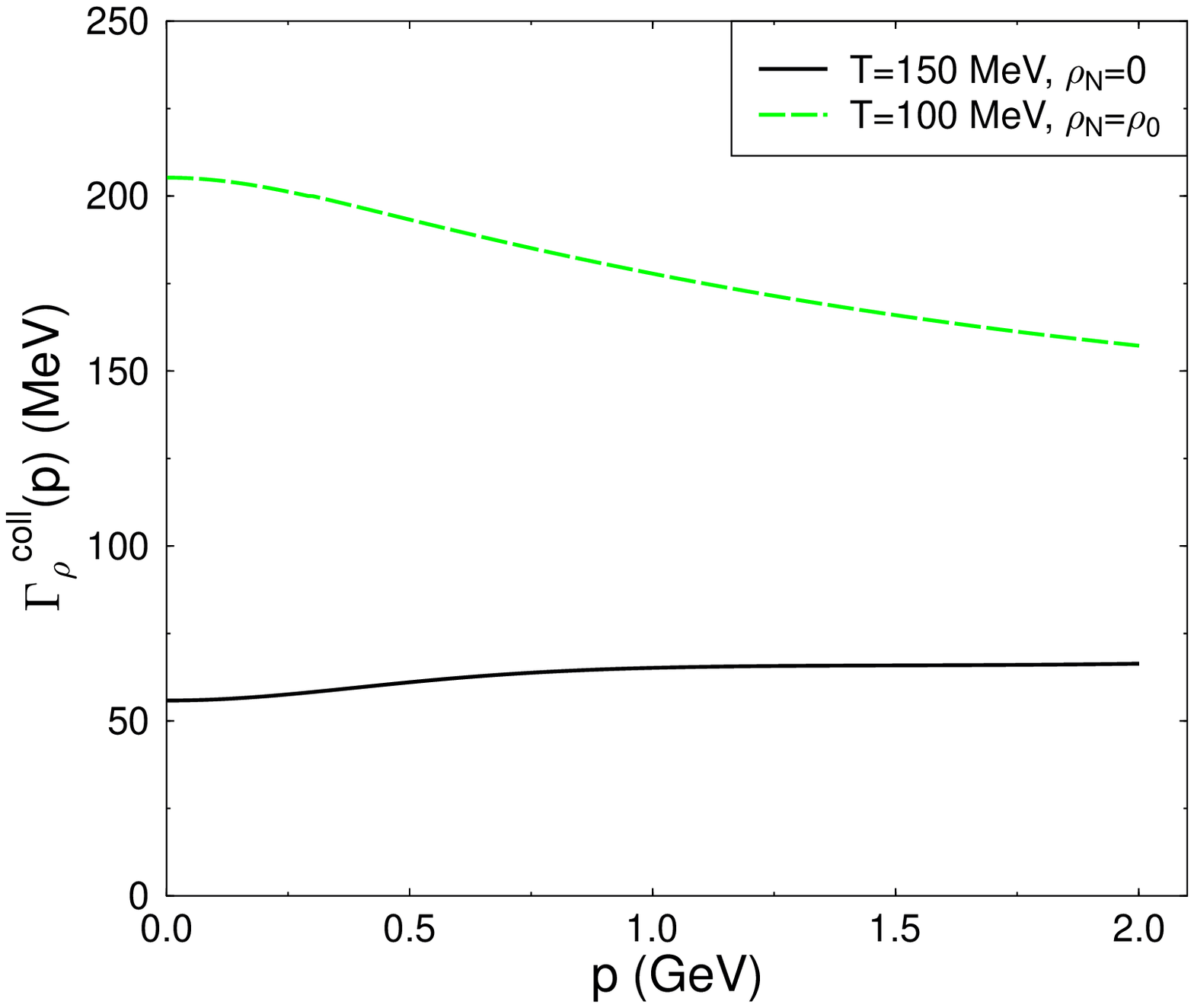,width=80mm}}} 
\vspace{-0.40cm}
\caption{$\rho$ collision rates in hadronic matter. Solid curve indicate
pure pion gas, dashed curve is in the medium of pion and nucleon.}
\label{fig:coll}
\end{minipage}
\hfill
\begin{minipage}[t]{70mm}
\vspace{-6.2cm}
\hspace{-0.4cm}
{\centerline{\epsfig{file=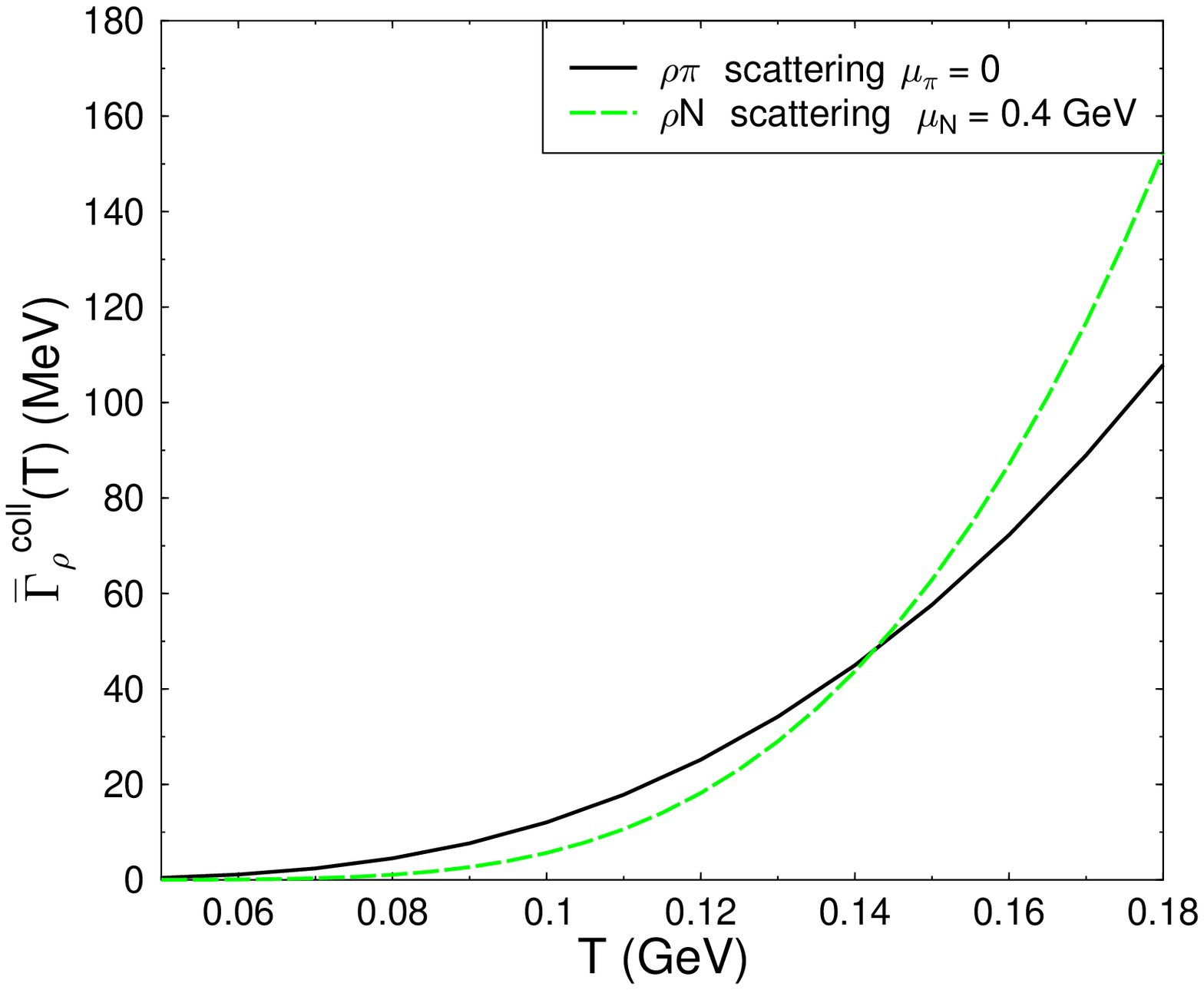,width=80mm}}}
\vspace{-0.40cm}
\caption{The thermal average of the $\rho$ collision rate $vs.$ the 
temperature. Solid curve  due to $\rho \, \pi$ interactions, dashed 
curve is $\rho$N collisions.}
\label{fig:collbar}
\end{minipage}
\end{figure}

The  elementary reactions which tend to
thermalize $\rho$'s in hot and dense matter are the channels
 $\rho_1\, \pi_2 \to \rho_3\, \pi_4$ and/or
$\rho_1\, N_2 \to \rho_3\, N_4$. A $\rho$ with arbitrary momentum $p_1$
(and energy $\omega$) can be captured by the thermal background.
A $\rho$ with momentum $p_1$ can also be produced from the thermal
background
by the inverse reaction: $\rho_3\, \pi_4 \to \rho_1\, \pi_2$. This inverse
rate is omitted in many similar treatments. The total collision rate for
bosons is the difference
$\Gamma^{\rm coll} (\omega) = \Gamma_d (\omega) - \Gamma_i (\omega)\, .$
We have explicitly verified that, for the physical
conditions relevant to this work, the inverse rate can be neglected, 
thus, the collision rate can be written as:
\be
\Gamma^{\rm coll} (\omega) \approx  \Gamma_d (\omega)
        = {1 \over {\omega}} \int \dpf{{\rm d}^3 p_2}{(2\pi)^3 2 E_2}\,
           {n}_2(E_2)\, \lambda^{1/2}(s, m_1^2, m_2^2)\, \sigma(s) \, ,
\ee
where $\sigma(s)$ is the cross section of $\rho \pi$ (or $\rho$N) collision.
But bear in mind that this practice can't however be
generalized to all cases and that at least the consideration of the
inverse rate remains important from the point of view of first
principles. 
Then we can calculate the collision rates of the $\rho$ in medium. 
For $\rho\, \pi$ scattering, we use effective
Lagrangians for the hadronic interaction \cite{gom,gg98},
the  scattering proceeds through s- and t-channels
respectively, where the $\pi,\, \rho, \, \omega, \, \phi, \, a_1$ and
$\omega'(1420)$ might be intermediate states, to calculated the cross
section which needed in the collision rate calculations \cite{kh95,ggesg}.
For $\rho$N scattering, we use the cross section from the Manley resonances
method\cite{kondratyuk}, furthermore, the resonances below the $\rho$ N
threshold, like the N$^*(1520)$ are also considered. In the rest frame of
the fluid, our results are shown in Fig. \ref{fig:coll}.

In the equilibrated matter, the thermal average of $\Gamma^{\rm coll}
(\omega)$ can be written as,
\be
{\overline \Gamma}^{\rm coll} =\dpf{{\cal N}_1 {\cal N}_2}{\rho_1}\,
        \int_{s_0}^{\infty} {\rm d} s \dpf{T}{2(2\pi)^4 {\sqrt s}}\,
        \lambda(s, m_1^2, m_2^2)\, {\tilde K}({\sqrt s}, m_1, m_2, T,
\mu_i)\, \sigma(s) \, ,
\ee
where $\rho_1$ is the $\rho$ meson density,
and $s_0=(m_1+m_2)^2$, when using Boltzmann statistics,
$ {\tilde K} = K_1({\sqrt s}/T) {\rm exp}[(\mu_1 +\mu_2)/T]$,  
with $K_1$ being the modified Bessel function.
Fig. \ref{fig:collbar}  shows the
the thermal average of the $\rho$ collision rates from $\rho \pi$ and 
$\rho$N collisions. 

\section{RESULTS AND CONCLUSION}

\begin{figure}[hbt]
\begin{minipage}[t]{70mm}
\vspace{-6.2cm}
{\centerline{\epsfig{file=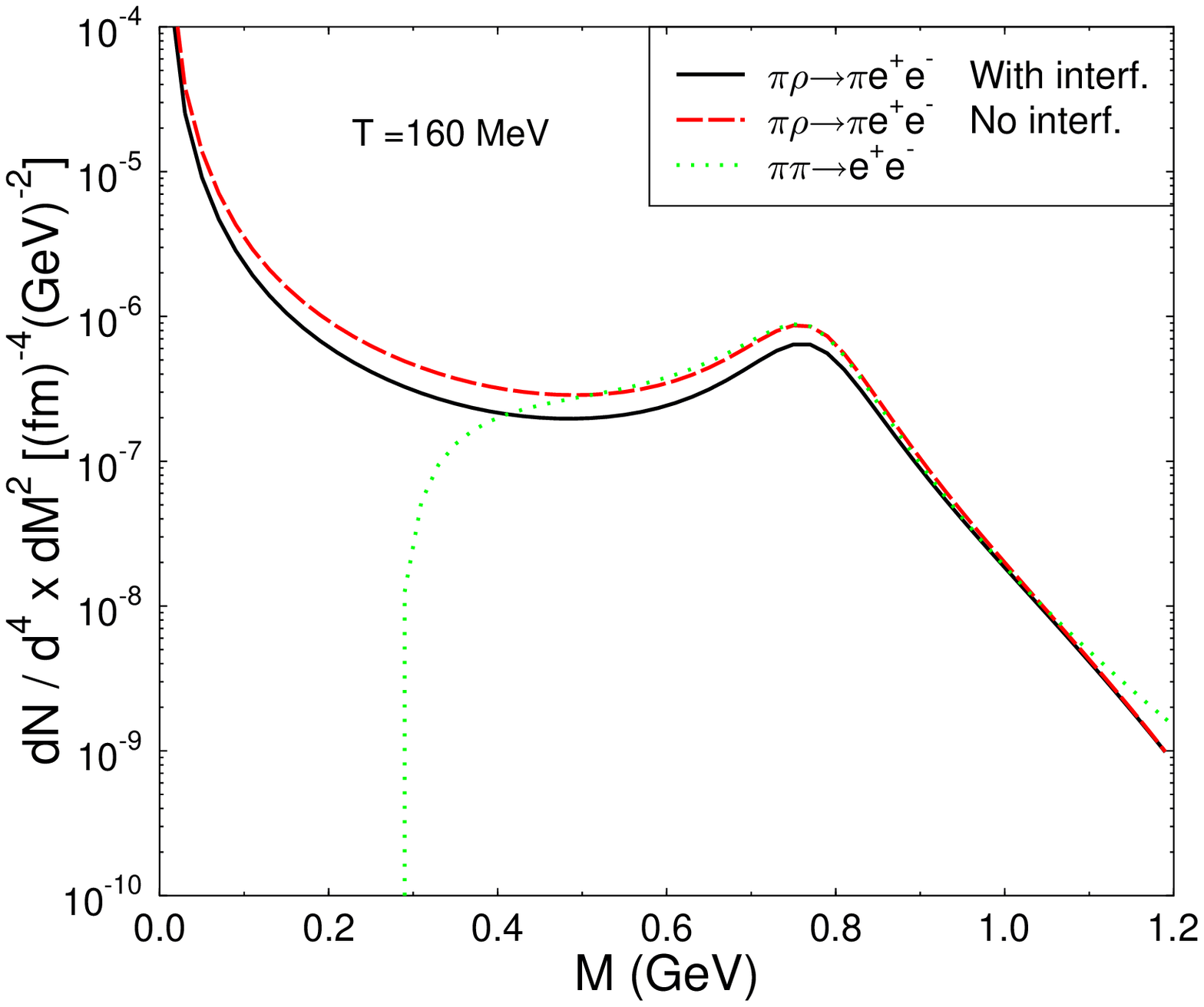,width=80mm}}} 
\vspace{-0.6cm}
\caption{Net rates for dielectrons, 
the $\rho$ has its vacuum width.}
\label{fig:ratefree}
\end{minipage}
\hfill
\begin{minipage}[t]{70mm}
\vspace{-6.2cm}
\hspace{-0.4cm}
{\centerline{\epsfig{file=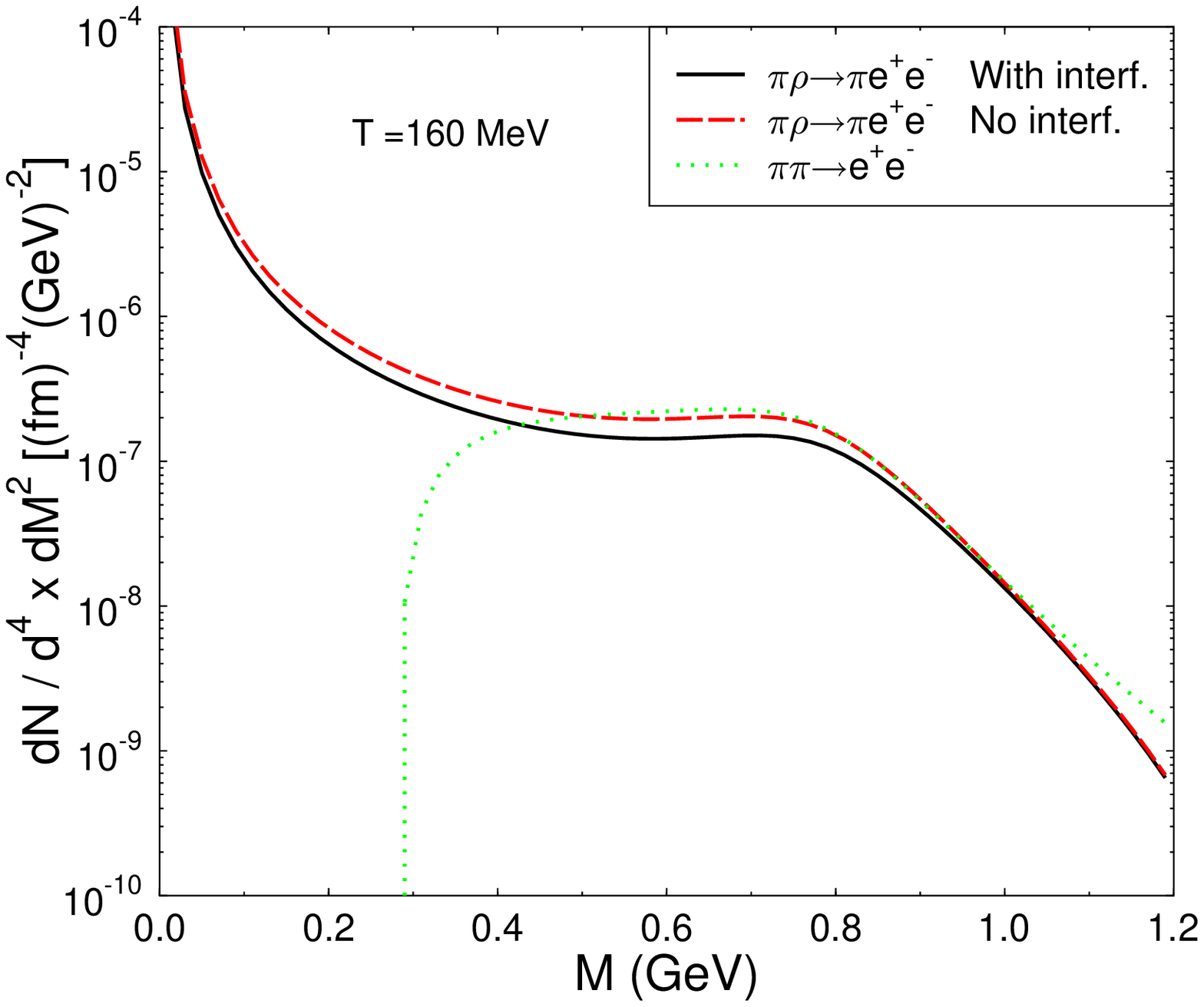,width=80mm}}} 
\vspace{-0.6cm}
\caption{Same as Fig. \protect\ref{fig:ratefree}, but the
$\rho$ has its collision-broadened width.}
\label{fig:ratecoll}
\end{minipage}
\end{figure}

In this section we will explore the results obtained by putting together the
different pieces we have described so far. In Fig. \ref{fig:ratefree}, 
we show the effects of the quantum interference resulting from coherently 
adding and then squaring the contributions with the same initial and final 
configuration. It follows from this plot that the quantum effects are not 
small and that their net effect is to decrease the dilepton signal in the 
mass region below the $\rho$. In Fig. \ref{fig:ratecoll},
we essentially repeat the calculations leading to the previous figure, with
one important exception: the rho width now has a temperature-dependent
component, owing to its interaction with other species in the
strongly-interacting thermal ensemble.

In summary, our results indicate that, for the collision rates, 
the contribution from
$\rho\,\pi$ collisions is the most important one in the high temperature
pion gas, while  at low temperatures and high density nuclear matter
the $\rho$N contribution is more important.
For dilepton production $\rho \pi \to  \pi e^+e^- $, the quantum 
interference effect
is to decrease the dilepton signal in the mass region below the $\rho$,
and due to temperature-dependent $\rho$ width, there is a large
suppression of the dilepton rates, especially around the $\rho$ peak.

\end{document}